\documentstyle[12pt]{article}
\newcommand{\bce}{\begin{center}}
\newcommand{\ece}{\end{center}}
\newcommand{\beq}{\begin{equation}}
\newcommand{\eeq}{\end{equation}}
\newcommand{\bea}{\vspace{0.25cm}\begin{eqnarray}}
\newcommand{\eea}{\end{eqnarray}}

\newcommand{\ba}{\begin{array}}
\newcommand{\ea}{\end{array}}

\newcommand{\doublespace}{
    \renewcommand{\baselinestretch}{1.6}\large\normalsize}

\def\lsim{\mathrel{\rlap{\lower4pt\hbox{\hskip1pt$\sim$}}
    \raise1pt\hbox{$<$}}}	  
\def\gsim{\mathrel{\rlap{\lower4pt\hbox{\hskip1pt$\sim$}}
    \raise1pt\hbox{$>$}}}	  

\def\Pom{{\bf I\!P}}

\def\lsim{\mathrel{\rlap{\lower4pt\hbox{\hskip1pt$\sim$}}
    \raise1pt\hbox{$<$}}}         
\def\gsim{\mathrel{\rlap{\lower4pt\hbox{\hskip1pt$\sim$}}
    \raise1pt\hbox{$>$}}}         

\def\Pom{{\bf I\!P}}

  \advance\oddsidemargin  by -1in
  \advance\evensidemargin by -1in
  \advance\topmargin      by -0.5in
\def\lsim{\mathrel{\rlap{\lower4pt\hbox{\hskip1pt$\sim$}}
    \raise1pt\hbox{$<$}}}         
\def\gsim{\mathrel{\rlap{\lower4pt\hbox{\hskip1pt$\sim$}}
    \raise1pt\hbox{$>$}}}         

\def\Pom{{\bf I\!P}}
\def\beq{\begin{equation}}
\def\endeq{\end{equation}}
\def\arr{\begin{eqnarray}}
\def\endarr{\end{eqnarray}}
\makeindex

\doublespace

\begin{document}


\phantom{.}{\bf \Large \hspace{10.0cm} KFA-IKP(Th)-1994-2 \\
\phantom{.}\hspace{11.9cm}4 January   1994\vspace{0.4cm}\\ }
\begin{center}
{\bf\sl \huge  The BFKL and GLDAP regimes  \\
for the perturbative QCD pomeron}
\vspace{0.4cm}\\
{\bf \large
N.N.~Nikolaev$^{a,b}$, B.G.~Zakharov$^{a,b}$ and V.R.Zoller$^{a,c}$
\bigskip\\}
{\it
$^{a}$IKP(Theorie), KFA J{\"u}lich, 5170 J{\"u}lich, Germany
\medskip\\
$^{b}$L. D. Landau Institute for Theoretical Physics, GSP-1,
117940, \\
ul. Kosygina 2, Moscow 117334, Russia.\medskip\\
$^{c}$ Institute for Theoretical and Experimental Physics,\\
Bolshaya Cheremushkinskaya 25, 117259 Moscow, Russia.
\vspace{1.0cm}\\ }
{\Large
Abstract}\\
\end{center}
The $S$-matrix of diffractive scattering is diagonalized in terms
of the colour dipole-dipole cross section.
Recently, we have shown that the dipole cross section satisfies
the generalized BFKL equation. In this paper we
discuss the spectrum and solutions of our
generalized BFKL
equation with allowance for the finite
gluon correlation radius $R_{c}$. The latter is introduced
in the gauge invariant manner. We present estimates of the
intercept of the pomeron and find the asymptotic form of the
dipole cross section. We consider the difference between the
BFKL and GLDAP evolutions and conclude that the GLDAP evolution
remains a viable description of deep inelastic scattering well
beyond the kinematical range of the HERA experiments.
We suggest methods of measuring the pomeron intercept in the
HERA experiments.
 \bigskip\\

\begin{center}
E-mail: kph154@zam001.zam.kfa-juelich.de
\end{center}

\pagebreak




\section{ Introduction.}


The asymptotic behavior of high-energy scattering in perturbative
QCD is usually discussed in terms of the Balitskii-Fadin-Kuraev-Lipatov
(BFKL) pomeron [1-3]. The BFKL equation is the integral equation
for the differential density of gluons, so that formulation of
predictions for the observable quantities like total cross section
is not straightforward. Furthermore, the original
BFKL equation is the scaling equation, and when discussing the
realistic QCD one encounters the difficult task of introducing
the infrared cutoff in the BFKL equation.
The scaling BFKL equation has a continuous spectrum and corresponds
to the vacuum singularity described by a cut in the complex
angular momentum plane. One of the pressing issues in the theory
of the perturbative QCD pomeron is whether the infrared cutoff
and the running QCD coupling can change the spectrum of the
pomeron and produce isolated poles.

In this paper we discuss the spectrum and solutions
of the generalized BFKL equation
directly for the total cross section, derived by us recently
 [4,5]. Our equation allows
to introduce a finite correlation radius for the perturbative
gluons in the gauge invariant manner.
The starting point
of its derivation is the technique of multiparton lightcone
wave functions, developed by two of the present authors [4,6].
The principal observation is that the
transverse separations $\vec{\rho}_{ij}=\vec{\rho}_{i}-
\vec{\rho}_{j}$ and
the lightcone momentum partitions $z_{i}$ of partons in the
many-body Fock state are conserved in the scattering
process. Interaction of the $(n+2)$-parton Fock state is
described by the lightcone wave function
$\Psi_{n+2}(\vec{\rho}_{n+2},z_{n+2},...,\vec{\rho}_{1},z_{1})$
and the $(n+2)$-parton cross section
$\sigma_{n+2}(\vec{\rho}_{n+2},...,\vec{\rho}_{1})$, which are
perturbatively calculable [4] (here $n$ referes to the number
of gluons in the Fock state).

To the lowest order in the perturbative  QCD  one starts
with the $q\bar{q}$ Fock states of mesons ($qqq$ state for the
baryons) and with the scattering of the ($A$)
projectile and ($B$) target colour dipoles of transverse size
$\vec{r}_{A}$ and $\vec{r}_{B}$ (here $\vec{r}_{A},\vec{r}_{B}$ are
the two-dimensional vectors in the impact parameter plane)
\beq
\sigma_{0}(\vec{r}_{A},\vec{r}_{B})=
{32 \over 9}
\int {d^{2}\vec{k}
\over(k^{2}+\mu_{G}^{2})^{2} }
\alpha_{S}^{2}
\left[1-\exp(-i\vec{k}\vec{r}_{A})\right]
\left[1-\exp(i\vec{k}\vec{r}_{B})\right]  \, .
\label{eq:1.1}
\endeq
Here
\beq
\alpha_{S}(k^{2})= {4\pi \over \beta_{0}
\log\left({k^{2} \over \Lambda_{QCD}^{2}}\right)}
\label{eq:1.2}
\endeq
is the running
strong coupling, $\beta_{0}=11-{2\over 3}N_{f}=9$ for $N_{f}=3$
active flavours, and in the integrand of the dipole-dipole
cross section $\alpha_{S}^{2}$ must be understood as
$\alpha_{S}({\rm max}\{k^{2},{C^{2}\over r_{A}^{2}}\})
\alpha_{S}({\rm max}\{k^{2},{C^{2}\over r_{B}^{2}}\})$, where
$C\approx 1.5$  [6]. Introduction of the infrared freezing of
the running coupling and
of the correlation radius for gluons $R_{c}=1/\mu_{G}$ is
discussed below.
In terms of the dipole-dipole cross section (\ref{eq:1.1}) the
perturbative part of the total cross section for the interaction
of mesons $A$ and $B$ equals
\beq
\sigma^{(pt)}(AB)=\langle \langle
\sigma(\vec{r}_{A},\vec{r}_{B}) \rangle_{A}
\rangle_{B}=
\int dz_{A} d^{2}\vec{r}_{A}
dz_{B} d^{2}\vec{r}_{B}
|\Psi(z_{A},\vec{r}_{A})|^{2}
|\Psi(z_{B},\vec{r}_{B})|^{2}
\sigma_{0}(\vec{r}_{A},\vec{r}_{B}) \,.
\label{eq:1.3}
\endeq
The irrefutable advantage of the representation
(\ref{eq:1.3}) is that it makes
full use of the exact
diagonalization of the scattering matrix in
the dipole-size representation.
(Hereafter we discuss $\sigma_{0}(\vec{r},\vec{R})$ averaged over
the relative orientation of dipoles, as it appears in
Eq.~(\ref{eq:1.3}).)
Notice the beam-target symmetry of the dipole-dipole cross
section,
\beq
\sigma_{0}(\vec{r}_{A},\vec{r}_{B})=
\sigma_{0}(\vec{r}_{B},\vec{r}_{A}) \, ,
\label{eq:1.4}
\endeq
and of the representation (\ref{eq:1.3}).

The increase of the perturbative component of the
total cross section
comes from the rising muliplicity of perturbative
gluons in hadrons, $n_{g} \propto \log s$, times
$\Delta\sigma_{g}$ - the change in the dipole cross
section for the presence of gluons:
$\Delta\sigma^{(pt)} \sim n_{g}\Delta\sigma_{g}$. (Here $s$ is
a square of the c.m.s. energy.)
The $n$-gluon Fock
components of the meson give contributions $\propto \log^{n}s$
to the total cross section. The crucial observation is that
the effects of higher order Fock states of the interacting hadrons
can be reabsorbed into the energy dependent dipole-dipole
cross section with retention of the representation (\ref{eq:1.3}).
This energy-dependent dipole cross section satisfies the
generalized BFKL equation derived in [4,5], investigation of
properties of which is the subject of the present paper. The
presentation is organized as follows:

In section 2 we briefly review the
derivation of our generalized BFKL equation for the dipole cross
and in section 3
discuss its BFKL scaling limit. In section 3 we also comment on
the introduction of the infrared cutoff into the scaling BFKL
equation. The subject of section 4 is the conventional QCD
evolution as the limiting case of our generalized BFKL equation.
The impact of
the finite correlation length for gluons and of the running QCD
coupling on the spectrum of the generalized BFKL equation is
discussed in detail in section 5. Our
principal conclusion is that the generalized BKL kernel has
a continuous spectrum, so that the partial waves of the
scattering amplitude have a cut in the complex angular momentum
plane. We also comment on the conditions under which the pomeron
can have a discret spectrum. In section 6 we find the
form of the dipole cross section for the
rightmost singularity in the $j$-plane and its intercept as a
function of the gluon correlation radius $R_{c}$. The subject of
section 7 is a transition from the GLDAP evolution to the BFKL
evolution with increasing energy. The beam-target symmetry of the
dipole-dipole cross section and
the admissible form of the boundary conditions
for the generalized BFKL equation are discussed in section 8.
In section 9 we comment on the restoration of the factorization
property of the asymptotic cross section. In section 10 we
suggest practical methods of measuring the intercept of the
pomeron in the HERA experiments on deep inelastic scattering.
In the Conclusions section
we summarize our basic results.

\section{Generalized BFKL equation for total cross section}

Now we sketch the derivation of our generalized BFKL equation [4,5]
for total cross sections. Unless specified otherwise, we consider
the contribution to the hadronic scattering from the exchange by
the perturbative gluons and suppress the superscript $(pt)$.
The perturbative $q\bar{q}g$ Fock state generated
radiatively from the parent colour-singlet $q\bar{q}$ state of
size $\vec{r}$ has the interaction cross section [4]
\beq
\sigma_{3}(r,\rho_{1},\rho_{2})=
{9 \over 8}[\sigma_{0}(\rho_{1})+\sigma_{0}(\rho_{2})] -
{1 \over 8}\sigma_{0}(r) \, ,
\label{eq:2.1}
\endeq
where $\vec{\rho}_{1,2}$ are separations of the
gluon from the quark and antiquark respectively, $\vec{\rho}_{2}=
\vec{\rho}_{1}-\vec{r}$. (Hereafter we suppress the target
variable $r_{B}$ and for the sake of brevity use $r=r_{A}$.)
The cross section $\sigma_{3}(r,\rho_{1},\rho_{2})$
has the gauge invariance
properties of $\sigma_{3}(r,0,r)=\sigma_{3}(r,r,0)=\sigma_{0}(r)$
and $\sigma_{3}(0,\rho,\rho)={9\over 4}\sigma_{0}(\rho)$ [4]. The
former shows that when the gluon is sitting on top of the (anti)quark,
the $qg (\bar{q}g)$ system is indistinguishable from the (anti)quark.
The latter shows that the colour-octet $q\bar{q}$ system of the
vanishing size is indistinguishable from the gluon, and ${9\over 4}$
is the familiar ratio of the octet and triplet couplings.
The increase of the cross section for the presence of
gluons equals
\beq
\Delta \sigma_{g}(r,\rho_{1},\rho_{2})=
\sigma_{3}(r,\rho_{1},\rho_{2}) -
\sigma_{0}(r)=
{9 \over 8}[\sigma_{0}(\rho_{1})+
\sigma_{0}(\rho_{2})-\sigma_{0}(r)]   \, \, ,
\label{eq:2.2}
\endeq

The lightcone density of soft, $z_{g}\ll 1$,
gluons in the $q\bar{q}g$ state derived
in [4] equals
\arr
|\Phi_{1}(\vec{r},\vec{\rho}_{1} ,\vec{\rho}_{2},z_{g})|^{2}=
{1 \over z_{g}} {1 \over 3\pi^{3}}
\mu_{G}^{2}
\left|g_{S}(r_{1}^{(min)})
K_{1}(\mu_{G}\rho_{1}){\vec{\rho}_{1}\over \rho_{1}}
-g_{S}(r_{2}^{(min)})
K_{1}(\mu_{G}\rho_{2}){\vec{\rho}_{2} \over \rho_{2}}\right|^{2} \, .
\label{eq:2.3}
\endarr
Here $g_{S}(r)$ is the running colour charge,
$\alpha_{S}(r)=g_{S}^{2}(r)/4\pi$ is the running strong coupling,
$r_{1,2}^{(min)}={\rm min}\{r,\rho_{1,2}\}$, $K_{1}(x)$ is the
modified Bessel function,
$z_{g}$ is a fraction of the (lightcone) momentum of
the $q\bar{q}$ pair carried by the gluon, and $\int dz_{g}/z_{g}
=log(s/s_{0})=\xi$ is the parameter of the Leading-$\log({1\over x})$
approximation and
gives the usual logarithmic multiplicity of
radiative gluons. The wave function (\ref{eq:2.3}) counts only
the physical, transverse, gluons [4].

If $n_{g}(r)$ is the number of perturbative gluons in the
dipole $\vec{r}$,
\beq
n_{g}(r)=
\int
dz_{g}\,d^{2}\vec{\rho}_{1}\,\,
|\Phi_{1}(\vec{r},\vec{\rho}_{1} ,\vec{\rho}_{2},z_{g})|^{2}\, ,
\label{eq:2.4}
\endeq
then the weight of the
radiationless $q\bar{q}$ component will be renormalized
by the factor $[1-n_{g}(z,\vec{r})]$. As a result, the total cross
section with allowance for the perturbative gluons in the beam
dipole $A$ takes the form
\arr
\sigma(\xi,r)=
[1-n_{g}(z,\vec{r})]\sigma_{0}(r)+
\int dz_{g}d^{2}\vec{\rho}\,\,
|\Phi_{1}(\vec{r},\vec{\rho}_{1} ,\vec{\rho}_{2},z_{g})|^{2}
\sigma_{3}(r,\rho_{1},\rho_{2})~~~~~~~~\nonumber\\
=
\sigma_{0}(r)+
\int dz_{g}d^{2}\vec{\rho}_{1}\,\,
|\Phi_{1}(\vec{r},\vec{\rho}_{1} ,\vec{\rho}_{2},z_{g})|^{2}
\Delta\sigma_{g}(r,\rho_{1},\rho_{2})=\sigma_{0}(r)+\sigma_{1}(r)\xi
=[1+\xi{\cal K}\otimes]\sigma_{0}(r)
 \,,
\label{eq:2.5}
\endarr
where the kernel ${\cal K}$ is defined by [4,5]
\arr
\sigma_{1}(r)={\cal K}\otimes\sigma_{0}(r)=~~~~~~~~~~~~~~~~~~~~~~~~~~~~
\nonumber\\
{3 \over 8\pi^{3}} \int d^{2}\vec{\rho}_{1}\,\,
\mu_{G}^{2}
\left|g_{S}(r_{1}^{(min)})
K_{1}(\mu_{G}\rho_{1}){\vec{\rho}_{1}\over \rho_{1}}
-g_{S}(r_{2}^{(min)})
K_{1}(\mu_{G}\rho_{2}){\vec{\rho}_{2} \over \rho_{2}}\right|^{2}
[\sigma_{0}(\rho_{1})+
\sigma_{0}(\rho_{2})-\sigma_{0}(r)]   \, \, .
\label{eq:2.6}
\endarr
Eq.~(\ref{eq:2.5}) shows that the effect of gluons can be reabsorbed
into the generalized, energy-dependent, dipole cross section
$\sigma(\xi,r)$.
To higher orders in $\xi$, this generalized dipole cross section
can be expanded as
$
\sigma(\xi,r)=\sum_{n=0}{1\over n!}\sigma_{n}(r)
\xi^{n} $,
where $\sigma_{n+1}={\cal K}\otimes \sigma_{n}$, so that
\beq
{\partial \sigma(\xi,r) \over \partial \xi} ={\cal K}\otimes
\sigma(\xi,r)
\label{eq:2.7}
\endeq
is our generalization of the BFKL equation for the dipole
cross section.

The colour gauge invariance of the presented formalism is noteworthy.
Firstly, the dipole-dipole cross section $\sigma_{0}(r_{A},r_{B})$
vanishes at $r_{A} \rightarrow 0$ or $r_{B}\rightarrow 0$, because
gluons decouple from the colour-singlet state of vanishing size.
Secondly, for the same reason the wave function (\ref{eq:2.3})
vanishes at $r\rightarrow 0$.
Thirdly,
$\Delta \sigma_{g}(r,\rho_{1},\rho_{2})\rightarrow 0$ when
$\rho_{1}\rightarrow 0$ (or $\rho_{2}\rightarrow 0$), since by
colour charge conservation the quark-gluon system with the gluon
sitting on top of the (anti)quark is indistinguishable from the
(anti)quark, and the interaction properties of such a $q\bar{q}g$
state are identical to that of the $q\bar{q}$ state.
Therefore, our introduction of the finite
correlation radius for gluons $R_{c}=1/\mu_{G}$, which
takes care of the perturbatie gluons not propagating beyond the
correlation radius $R_{c}$, is perfectly consistent
with the gauge invariance.
For instance, Eq.~(\ref{eq:2.6}) supports the gauge invariance
constraint $\sigma_{n}(r)\rightarrow 0$ at $r\rightarrow 0$, to
all orders $n$.
(We do not have a complete proof of the gauge invariance, though.
Neither can we prove that our prescription is unique.)

The renormalization of the weight of the radiationless $q\bar{q}$
Fock state in Eq.~(\ref{eq:2.5})
in a simple and intuitively appealing form takes care of the
virtual radiative corrections (in the BFKL formalism [1,2]
these very radiative corrections
are responsible for the reggeization of gluons).
Notice, that although the multiplicity of radiative gluons
$n_{g}(r)$, which appeared in the intermediate
stage of the derivation of our Eq.~(\ref{eq:2.6}), is formally
divergent, the generalized dipole cross section $\sigma(\xi,r)$
and Eqs.~(\ref{eq:2.6},\ref{eq:2.7}) are both ultraviolet and
infrared finite.



\section{The BFKL scaling limit}

In the BFKL scaling limit of $r, \rho_{1},\rho_{2} \ll R_{c}$
and of the fixed $\alpha_{S}$
\beq
\mu_{G}^{2}\left|K_{1}(\mu_{G}\rho_{1}){\vec{\rho}_{1}\over \rho_{1}}
-K_{1}(\mu_{G}\rho_{2}){\vec{\rho}_{2}\over\rho_{2}}\right|^{2} =
{r^{2} \over \rho_{1}^{2}\rho_{2}^{2}}   \, ,
\label{eq:3.1}
\endeq
the kernel ${\cal K}$ becomes independent of the gluon correlation
radius $R_{c}$ and with the fixed $\alpha_{S}$ it takes on the
scale-invariant form. The corresponding
eigenfunctions of Eq.~(\ref{eq:2.7}) are
\beq
E(\omega,\xi,r)=(r^{2})^{{1\over 2}+\omega}
\exp[\xi\Delta(\omega)]
\label{eq:3.2}
\endeq
with the eigenvalue (intercept)
[here $\vec{r}=r\vec{n}$,  $\vec{\rho}_{1}=r\vec{x}$  and
$\vec{\rho}_{2}=r(\vec{x}+\vec{n})$]
\arr
\Delta(\omega) = \lim_{\epsilon \to 0}
{3\alpha_{S} \over 2\pi^{2}}
\int d^{2}\vec{x}~~{2(\vec{x}^{2})^{{1\over 2}+\omega}-1
\over [\vec{x}^{2} +\epsilon^{2}]
[(\vec{x}+\vec{n})^{2}+\epsilon^{2}]}=
{3\alpha_{S} \over \pi}
\int_{0}^{1} dz
{z^{{1\over 2}-\omega}+z^{{1\over 2}+\omega}-2z \over z(1-z)}
\nonumber\\
=
{3\alpha_{S} \over \pi} [2\Psi(1)-
\Psi({1\over 2}-\omega)-\Psi({1\over 2}+\omega)]\, . ~~~~~~~~~~~~~~
\label{eq:3.3}
\endarr
Here $\Psi(x)$ is the digamma function, and
we have indicated the regularization which preserves the
symmetry of the kernel ${\cal K}$.
The final result for $\Delta(\omega)$ coincides with eigenvalues of
the BFKL equation found in [1-3]. The solution (\ref{eq:3.2})
corresponds to the
singularity in the complex-$j$ plane, located at $j=1+\Delta(\omega)$.
The rightmost singularity
is located at $j=\alpha_{\Pom}=1+\Delta_{\Pom}$, where
\beq
\Delta_{\Pom}=\Delta(0)=
{12\log2\over \pi}\alpha_{S} \, .
\label{eq:3.4}
\endeq

When $\omega$ is real and varies from $-{1\over 2}$ to
0 and to ${1\over 2}$, also the intercept
$\Delta(\omega)$ is real and varies from $+\infty$ down to
$\Delta(0)=\Delta_{\Pom}$ and back to $+\infty$, along the cut
from $j=1+\Delta_{\Pom}$ to $+\infty$ in
the complex angular momentum $j$ plane. If $\omega =i\nu$ and $\nu$
varies from $-\infty$ to
0 and to $+\infty$, then the intercept
$\Delta(i\nu)$ is again real and varies from $-\infty$ up to
$\Delta(0)=\Delta_{\Pom}$ and back to $-\infty$, along the cut
from $j=-\infty$ to $j=1+\Delta_{\Pom}$
in the complex $j$-plane. The choice of the latter cut
is appropriate for the Regge asymptotics at $\xi \gg 1$ and
the counterpart of the conventional Mellin representation is
\beq
\sigma(\xi,r)=
\int_{-\infty}^{+\infty} d\nu \,f(\nu)E(i\nu,r,\xi) =
r\int_{-\infty}^{+\infty} d\nu
f(\nu)\exp[2i\nu\log(r)]\exp(\Delta(i\nu)\xi) \, .
\label{eq:3.5}
\endeq
The Mellin transform is obtained from Eq.~(\ref{eq:3.5}) if one
changes the integration variable from the $\nu$ to the angular
momentum $j=1+\Delta(i\nu)$.
The spectral amplitude $f(\nu)$ is determined by the
boundary condition $\sigma(\xi=0,r)$:
\beq
f(\nu)={1\over \pi}\int dr{\sigma(0,r)\over r^{2}}
\exp[-2i\nu\log(r)] \,\,.
\label{eq:3.6}
\endeq
Eqs.~(\ref{eq:3.5},\ref{eq:3.6}) give a nontrivial connection between
the $r$-dependence of the total cross section and its energy
dependence. In the BFKL scaling
regime, the rightmost $j$-plane singularity
corresponds to the asymptotic dipole cross section
$\sigma_{\Pom}(\xi,r) = \sigma_{\Pom}(r)\exp(\xi\Delta_{\Pom})$,
where
\beq
\sigma_{\Pom}(r) \propto r \, .
\label{eq:3.7}
\endeq

More direct correspondance between our equations
(\ref{eq:2.6},\ref{eq:2.7}) in the scaling limit of
$\mu_{G}\rightarrow 0$ and the conventional
BFKL equation can be established
if one rewrites Eqs.~(\ref{eq:2.6},\ref{eq:2.7}) as an equation for
the function
\beq
g(\xi,r)={3 \sigma(\xi,r)\over \pi^{2}\alpha_{S}r^{2}}\,
\label{eq:3.8}
\endeq
which in the BFKL scaling limit is simply the density of gluons
$g(x,k^{2})$
at the Bjorken variable $x=x_{0}\exp(-\xi)$ and the virtuality
$k^{2} \sim 1/r^{2}$, where $x_{0}\sim $0.1-0.01 corresponds to
the onset of the leading-$\log({1\over x})$ approximation
 (the relation (\ref{eq:3.8}) holds also
for the
running strong coupling $\alpha_{S}=\alpha_{S}(r)$,
for the detailed derivation see [4,7]).
The transformation from the size-$\vec{r}$ representation
to the momentum-$\vec{p}$ representation can easily be performed
making use of the conformal symmetry of the BFKL
equation [3]. Namely, by virtue of the conformal symmetry
$\vec{r}$ can easily be traded for $\vec{p}$
and $\sigma(\xi,r)$ can be traded for $g(\xi,p^{2})$ with the
result
\arr
{\partial g(\xi,p^{2})\over \partial\xi}=
{3\alpha_{S} \over 2\pi^{2}} \int d^{2}\vec{k}\,\,
{\vec{p}\,^{2}
\over \vec{k}^{2}(\vec{p}-\vec{k})^{2}}
\left[
g(\xi,k^{2})+
g(\xi,(\vec{p}-\vec{k})^{2})-
g(\xi,p^{2})\right]
 \, \, .
\label{eq:3.9}
\endarr
The factor $p^{2}$ in the integrand of (\ref{eq:3.9}) is crucial
for supporting the gauge invariance constraint $g(\xi,p^{2})
\rightarrow 0$ at $p^{2}\rightarrow 0$, which is the counterpart
of the gauge invariance constraint $\sigma(\xi,r) \rightarrow 0$
at $r\rightarrow 0$ in the $\vec{r}$-representation. If one introduces
$\phi(\xi,k^{2})=g(\xi,k^{2})/k^{2}$, then
the original form of the BFKL equation
will be recovered from Eq.~(\ref{eq:3.9}) (in the scaling limit
$\phi(\xi,k^{2})$ satisfies the same equation as the more often
considered $\partial g(\xi,k^{2})/\partial k^{2}$).

Introduction of the infrared regularization into the BFKL scaling
equation (\ref{eq:3.9}) is one of the pressing issues in the
theory of the perturbative QCD pomeron. We are going to use
neither Eq.~(\ref{eq:3.9}) nor any of its modifications to be
discussed below; we resort to our Eqs.~(\ref{eq:2.6},\ref{eq:2.7}),
which allow to treat effects of both the gluon correlation
radius $R_{c}$ and the running strong coupling in the consistent
and manifestly gauge invariant manner. Nonetheless, in order to
establish a connection with the existing literature, we comment
on the infrared regularization of the BFKL equation (\ref{eq:3.9})
in the momentum representation.

Our form (\ref{eq:3.9}) of the BFKL equation
is more convenient for these purposes, as it allows the clearcut
separation of propagators $1/\vec{k}^{2}$ and
$1/(\vec{p}-\vec{k})^{2}$ from the factor $\vec{p}\,^{2}$ in the
nominator in Eq.~(\ref{eq:3.9}). The latter, as well as the
factor $r^{2}$ in Eq.~(\ref{eq:3.1}), has its
origin in cancellations of the
radiation of soft gluons by the colour singlet states. Therefore,
the infrared regularization of Eq.~(\ref{eq:3.9}) which plausibly
respects the gauge invariance, is the substitution
\beq
{\vec{p}\,^{2}
\over \vec{k}^{2}(\vec{p}-\vec{k})^{2}}  \Longrightarrow
{\vec{p}\,^{2}
\over [\vec{k}^{2}+m^{2}][(\vec{p}-\vec{k})^{2}+m^{2}]} \,,
\label{eq:3.10}
\endeq
which treats the both gluon propagators on the equal footing.
In the scaling limit of $p^{2} \gg m^{2}$,
the function $\psi(\xi,p^{2})=g(\xi,p^{2})/(p^{2}+m^{2})$
is similar to, and satisfies the same equation as,
$\partial g(\xi,p^{2})/\partial p^{2}$.
The equation for $\psi(\xi,p^{2})$, which follows from (\ref{eq:3.9})
subject to the substitution (\ref{eq:3.10}), reads
\arr
{\partial \psi(\xi,p^{2})\over \partial\xi}=
{3\alpha_{S} \over 2\pi^{2}}\cdot
{\vec{p}\,^{2} \over
\vec{p}\,^{2}+m^{2}}\,
\int  d^{2}\vec{k}\,
\left[
{\psi(\xi,k^{2})\over (\vec{p}-\vec{k})^{2}+m^{2}}
+{\psi(\xi,(\vec{p}-\vec{k})^{2})\over \vec{k}^{2}+m^{2}}
-{(p^{2}+m^{2})\psi(\xi,p^{2})\over[(\vec{p}-\vec{k})^{2}+m^{2}]
[\vec{k}^{2}+m^{2}]}\right] \nonumber \\
={3\alpha_{S} \over \pi^{2}} \cdot
{\vec{p}\,^{2} \over
\vec{p}\,^{2}+m^{2}}\,
\int  d^{2}\vec{k}\,
\left[
{\psi(\xi,k^{2})\over (\vec{p}-\vec{k})^{2}+m^{2}}
-{(p^{2}+m^{2})\psi(\xi,p^{2})\over[(\vec{p}-\vec{k})^{2}+m^{2}]
[\vec{k}^{2}+(\vec{p}-\vec{k})^{2}+2m^{2}]}\right]
\label{eq:3.11}
\endarr
The asymmetric form in the second line of Eq.~(\ref{eq:3.11})
is obtained from the symmetric form in the first line of
Eq.~(\ref{eq:3.11}) if one uses the identity
[2]
\beq
\int {d^{2}\vec{k}\over
[\vec{k}^{2}+m^{2}]
[(\vec{p}-\vec{k})^{2}+m^{2}] }
=
2\int{d^{2}\vec{k}\over
[\vec{k}^{2}+m^{2}]
[(\vec{p}-\vec{k})^{2}+\vec{k}^{2}+2m^{2}] } \, .
\label{eq:3.12}
\endeq
(Erroneous form of this identity is cited in [8]).
The asymmetric form of Eq.~(\ref{eq:3.11}) at $m^{2}=0$ is
precisely the Balitskii-Lipatov form [2] of the BFKL equation,
which is the one customarily used in the recent literature.

One often imposes on this asymmetric version of the BFKL equation
the infrared cutoff $k^{2}\geq k_{0}^{2}$ [9-11], which
is illegitimate. Indeed, such a cutoff does not ensure, and
as matter of fact manifestly breaks, the condition of the
identical infrared cutoff of different
gluon propagators, as it breaks the initial symmetry property of
the BFKL equation (\ref{eq:3.9}) (for a similar criticism of the
Collins-Kwiecinski (CK) sharp cutoff
$k^{2}\geq k_{0}^{2}$ [9] see also
Collins and Landshoff [10]). Furthermore, such a cutoff is
evidently suspect from the point of view of the gauge invariance.
One must rather use
the equation (\ref{eq:3.11}) which introduces the infrared
regularization with retention of the symmetric cutoff of the
gluon propagators and in the manifestly gauge invariant manner.
Balitskii and Lipatov [2] did not encounter these problems, as
they introduce the massive vector mesons in a consistent manner,
using the spontaneous
breaking of gauge symmetry.
Introduction of massive gluons into
the BFKL equation was also discussed by Ross and Hancock [11].
The Ross-Hancock prescription is
different from ours in Eq.~(\ref{eq:3.11}),
manifestly breaks the symmetry of the cutoff of
gluon propagators and,
in contrast to our Eq.~(\ref{eq:3.11}), does not support the
$\vec{p}\,^{2} \rightarrow 0$ gauge invariance constraints.



\section{The GLDAP limit}

The limit of large but finite
$\xi$ and very small $r^{2}\ll R^{2}$ is often referred to as the
Double-Leading-Logarithmic Approximation (DLLA).
Here $R$ is either the target size, $R\sim r_{B}$ or the
correlation radius $R_{c}$ if $R_{c} < r_{B}$.
This is the limit typical of the deep inelastic scattering, and
it is usually described by the GLDAP evolution equations [12].
Introduction of the gluon correlation radius $R_{c}$
and of the running QCD coupling
poses no problems in the DLLA. In the DLLA,
$\xi=\log({x_{0}\over x})$, where $x$ is the Bjorken variable,
and $x_{0}\lsim 0.1-0.01$ is the threshold for the
Leading-$\log({1\over x})$ approximation.
The BFKL equation contains the GLDAP equation as the limiting case.
In the DLLA the kernel ${\cal K}$ takes a particularly simple
form [4]
\beq
\sigma_{n+1}(r)={\cal K}\otimes \sigma_{n}(r) =
{3 r^{2}
\alpha_{S}(r)
\over \pi^{2}}\int_{r^{2}}^{R^{2}}{d^{2}\vec{\rho}
\over \rho^{4}}
\sigma_{n}(\rho)
\, ,
\label{eq:4.1}
\endeq
which is equivalent to the GLDAP evolution equation. The dominant
contribution comes from the DLLA ordering of sizes
$r^{2}\ll \rho^{2}\ll R^{2}$, which justifies factoring out
$\alpha_{S}(r)$ in Eq.~(\ref{eq:4.1}).
 In the same DLLA the boundary
condition, the dipole-dipole cross section (\ref{eq:1.3}), equals
\beq
\sigma(\xi=0,r)=
\sigma_{0}(r,R) \approx Cr^{2}\alpha_{S}(r)L(R,r)\, ,
\label{eq:4.2}
\endeq
where $L(R,r)$ is the parameter of the Leading-$\log(r^{2})$
approximation (the Leading-$\log(Q^{2})$ approximation in the
more conventional momentum representation),
\beq
L(R,r)\approx \log\left[\alpha_{S}(R)\over \alpha_{S}(r)\right] \, ,
\label{eq:4.3}
\endeq
Iterations of Eq.~(\ref{eq:4.1}) subject to the boundary condition
(\ref{eq:4.2}) give the DLLA solution [4]
\beq
\sigma_{n+1}(r)={\cal K}\otimes \sigma_{n}(r) =
{1\over n+1}\cdot{12\over \beta_{0}}L(R,r)\sigma_{n}(r)\, .
\label{eq:4.4}
\endeq
The corresponding generalized dipole cross section has the DLLA
asymptotics
\beq
\sigma_{DLLA}(\xi,r)=\sigma(0,r)\sum_{n=0}{\eta^{n} \over n!(n+1)!}
\sim r^{2}\alpha_{S}(r)L(r,R){\exp(2\sqrt{\eta}) \over \sqrt{\eta}} \, ,
\label{eq:4.5}
\endeq
where $\eta$ is the expansion parameter of DLLA, which is a product
of the leading-$\log(s)$ and the leading-$\log(r^{2})$ expansion
parameters:
\beq
\eta ={12\over \beta_{0}}\xi L(r,R)
={4\over 3}\xi L(r,R)\, .
\label{eq:4.6}
\endeq
Notice, that the DLLA solution implicitly invovles an
assumption that at large $r\sim R$, the cross section
$\sigma_{DLLA}(\xi,r)$ is approximately constant.

The large-$n$ behavior of the DLLA iterations (\ref{eq:4.4}) must
be compared to
\beq
\sigma_{n+1}(r)=\Delta_{\Pom}\sigma_{n}(r)
\label{eq:4.7}
\endeq
for the leading BFKL solution, which suggests the criterion for
the breaking of the DLLA and/or GLDAP evolution: At large $\eta$
the DLLA cross section (\ref{eq:4.5}) is dominated by the
contribution from large $n \sim \sqrt{\eta}$. The DLLA to
GLDAP evolution
breaks when
\beq
{L(R,r)\over n} \approx {L(R,r)\over \sqrt{\eta}} \lsim
{3\over 4} \Delta_{\Pom}\,.
\label{eq:4.8}
\endeq
The pattern of this DLLA breaking in the case of the running
$\alpha_{S}(r)$ will be discussed in more detail below. Here we
only wish to emphasize that in the physically interesting case
of the running coupling and of finite correlation radius for the
perturbative gluons $R_{c}$ the intercept $\Delta_{\Pom}$ is the
constant, which does not depend on $r$. Therefore, the boundary
line $x=x_{c}(r)$ in the $(r,\xi)$ plane
between the BFKL evolution
and the GLDAP evolution to DLLA
is given by
\beq
\log\left({x_{0}\over x_{c}(r)}\right)=
{4\over 3\Delta_{\Pom}^{2}}\log\left[{1\over \alpha_{S}(r)}\right] \, .
\label{eq:4.9}
\endeq
It is different from the much discussed erroneous boundary
suggested in [8]. We shall comment more on the transition from
the DLLA to GLDAP evolution to the BFKL evolution
below, now we concentrate on the investigation of the spectrum
and eigenfunctions of our generalized BFKL equation.



\section{The spectrum and eigenfunctions of the generalized
BFKL equation}



\subsection{The Green's function, diffusion
and the 'Schr\"odinger' equation}

The generic solution of Eq.~(\ref{eq:2.7}) can be written as
\beq
\sigma(\xi,r)=r\int {dr'\over (r')^{2}}K(\xi,r,r')\sigma(\xi=0,r')\, ,
\label{eq:5.1.1}
\endeq
where in the BFKL regime the evolution kernel (Green's function)
equals
\beq
K(\xi,r,r')={1\over \pi}\int d\nu \exp\left[2i\nu\log{r\over r'}
\right]\exp[\xi \Delta(i\nu)] \, .
\label{eq:5.1.2}
\endeq
Since $\Delta(i\nu)$ has the maximum
at $\nu=0$,
\beq
\Delta(i\nu)=\Delta_{\Pom}-{1\over 2}\Delta''(0)\nu^{2} \, ,
\label{eq:5.1.3}
\endeq
the large-$\xi$ behavior of the Green's function can be evaluated
in the saddle-point approximation:
\beq
K(\xi,r,r')\propto
{\exp(\Delta_{\Pom}\xi) \over \sqrt{\xi}}
\exp\left[- {(\log r^{2}-\log (r')^{2})^{2}\over
2\xi \Delta^{''}(0)}\right] \, .
\label{eq:5.1.4}
\endeq
Since the BFKL eigenfunctions (\ref{eq:3.2})
are oscillating functions of $r$, {\sl\'a priori}
it is not obvious that an arbitrary solution (\ref{eq:3.5}) will be
positive-valued at all $\xi$ and $r$.
The BFKL
kernel $K(\xi,r,r')$ of Eq.~(\ref{eq:5.1.2}) is manifestly
positive-valued at
large $\xi$, so that starting with the positive-valued
$\sigma(0,r)$ one obtaines the positive-valued asymptotical cross
section $\sigma(\xi,r)$.

The `diffusion' kernel (\ref{eq:5.1.2}) makes it obvious that
the BFKL scaling approximation is not self-consistent in the
realm of realistic QCD:
starting with $\sigma(\xi=0,r)$ which was concentrated at the
perturbative small $r \lsim R \ll R_{c}$ one ends up at large
$\xi$ with $\sigma(\xi,r)$ which extends up to the
nonperturbative $r\sim R\exp(\sqrt{\xi\Delta''(0)}) > R_{c}$ .
Therefore, introduction of a certain infrared regularization
in the form of a finite gluon correlation length $R_{c}$ is
inevitable, and hereafter we concentrate on effects of finite
$R_{c}$ on the intercept $\Delta_{\Pom}$.
Interpretation of $\mu_{G}=1/R_{c}$ as an effective
mass of the gluon suggests the infrared freezing of the strong
coupling $\alpha_{S}(r)$ at $r> R_{f} \sim R_{c}$.
In the numerical analysis we use the running coupling
\beq
\alpha_{S}(r)={4\pi \over
\beta_{0}\log\left(
{C^{2} \over \Lambda_{QCD}^{2}r^{2}}\right)} \, ,
\label{eq:5.1.5}
\endeq
where $C\approx 1.5$ [6]. At large $r$ we impose the simplest
freezing $\alpha_{S}(r>R_{f})=\alpha_{S}^{(fr)}=0.8$. This
corresponds to the freezing radius $R_{f}\approx 0.42$f and/or
the freezing momentum $k_{f}\approx 0.72\,{\rm GeV/c}$ in the
momentum representation (\ref{eq:1.2}).
The more sophisticated smooth freezing can easily be considered, but
it will be obvious that our principal conclusions do not depend
on the form of the freezing.
The gluon correlation radius $R_{c}$ and the freezing coupling
provide the minimal infrared regularization of the perturbation
theory, and in the sequel we consider the so-regularized
generalized BFKL equation (\ref{eq:2.6},\ref{eq:2.7}) as applicable
at both small and large radii $r$.

Although only the case of $R_{c}\approx R_{f}$ is of the physical
interest, a study of (albeit unphysical)
limiting cases $R_{f}\rightarrow 0$ at finite $R_{c}$, and of
finite $R_{f}$ at $R_{c}\rightarrow \infty$,
is instructive for the insight into
how the spectrum of the $j$-plane singularities is modified by
the infrared regularization. The corresponding analysis is greatly
facilitated by the observation, that in view of
Eqs.~(\ref{eq:5.1.1},\ref{eq:5.1.4}) the large-$\xi$ behavior of
solutions of the BFKL limit of  Eq.~(\ref{eq:2.7}) is similar
to the large-$\xi$ behavior of solutions of the
'Schr\"odinger' equation
\beq
\left[-{\Delta''(0)\over 2}{\partial^{2} \over \partial z^{2} }
+V(z)\right]\Phi= -{\partial \over \partial \xi} \Phi
=\epsilon \Phi
\label{eq:5.1.6}
\endeq
for a particle of mass  $M=1/\Delta''(0)$ in the potential
$V(z)=-\Delta(0)$. Here the coordinate $z=\log r^{2}$,
the 'wavefunction' $\Phi(z,\xi)=\sigma(\xi,r)/r$ and the
intercept equals the 'energy' $\epsilon$
taken with the minus sign.


\subsection{Finite correlation radius $R_{c}$, fixed coupling
$\alpha_{S}$}

The first limiting case of $R_{f}\rightarrow 0$ corresponds to
introduction of a  finite gluon correlation radius $R_{c}$
at fixed $\alpha_{S}$. In this limiting case the intercept
$\Delta_{\Pom}$ will still be given by the BFKL formula (\ref{eq:3.4}).
Indeed, on the infinite semiaxis $\log r < \log R_{c}$
the kernel ${\cal K}$ retains its scaling properties,
the corresponding eigenfunctions  will be essentially identical to
the set (\ref{eq:3.2}), the spectrum of eigenvalues will be continuous
and the cut in the $j$-plane will be the same as
at $R_{c} \rightarrow \infty$. This is particularly obvious from
the Schr\"odinger Eq.~(\ref{eq:5.1.6}), since on
the the semiaxis $z<z_{c}=\log R_{c}^{2}$
the potential $V(z)=-\Delta_{\Pom}$ is flat, and
the corresponding Schr\"odinger operator (\ref{eq:5.1.6})
has a continuous spectrum starting with the minimal energy
$\epsilon = -\Delta_{\Pom}$.
Evidently, this property of the spectrum of the
Schr\"odinger operator (\ref{eq:5.1.6})
does not depend on the details of how the gluon correlation length
$R_{c}$ is introduced. This is a reason why the
similar conclusion on the spectrum of the infrared-cutoff BFKL
equation was reached in [9] in a model with the above criticized
sharp $k^{2}\geq k_{0}^{2}$ cutoff in the momentum representation.

The behaviour of solutions at large $r\gg R_{c}$
requires special investigation. In this region the term
$\propto K_{1}(\mu_{G}\rho_{1})K_{1}(\mu_{G}\rho_{2})$ in the
kernel ${\cal K}$ is exponentially
small, which is related to the exponential decay of the correlation
function (the propagator) of perturbative gluons. (Such an
exponential decay of the gluon correlation function with the
correlation radius $R_{c}=0.2-0.4$f is suggested by the lattice
QCD studies, for the review see [13]. We remind that $K_{1}(x)$
in the kernel ${\cal K}$ comes from the gradient of the gluon
correlation function $\propto K_{0}(x)$). Then,
at large $r$ the kernel ${\cal K}$ of our Eq.~(\ref{eq:2.7})
will be dominated by the contributions from $\rho_{1}\lsim R_{c}
\ll \rho_{2} \approx r$ and from
$\rho_{2}\lsim R_{c} \ll \rho_{1}\approx r$, will have the limiting
value which does not depend on $r$, and
Eq.~(\ref{eq:2.7}) takes the form
\arr
\sigma_{n+1}(r)=
{3\alpha_{S} \over \pi^{3}} \int d^{2}\vec{\rho}_{1}\,\,
\mu_{G}^{2}
K_{1}(\mu_{G}\rho_{1})^{2}
[\sigma_{n}(\rho_{1}) +\sigma_{n}(\rho_{2})-\sigma_{n}(r)]\, .
\label{eq:5.2.1}
\endarr
This equation  has a continuum of
solutions with the large-$r$ behavior of the form
\beq
E_{+}(\beta,\xi,r)= [a(\beta)+\exp(i\beta\mu_{G}r)]
\exp[\delta(\beta)\xi]\, .
\label{eq:5.2.2}
\endeq
(Apparently, one can put $a(\beta)=0$, but this is unimportant
for the subsequent analysis).

When continued to small $r$
through the region of $r\sim R_{c}$, the plane waves
in the linear-$r$ space transform into the plane waves in the
$\log r^{2}$-space (times the overall factor $r$),
$E(i\nu,\xi,r)=r\exp(i\nu\log r^{2})$ of Eq.~(\ref{eq:3.2}),
so that $\delta(\beta)=\Delta(i\nu)$.
Evidently, the two real and the node-free
solutions with $\nu=0$ and $\beta=0$ must match each
other, so that the rightmost $j$-plane singularity with the
intercept $\Delta_{\Pom}$ Eq.~(\ref{eq:3.4})
must correspond to $\sigma_{\Pom}(r)\propto r$ at $r \ll R_{c}$ and
$\sigma_{\Pom}(r)\propto const$ at $r \gg R_{c}$.
We wish to emphasize that although such a $\sigma_{\Pom}(r)$ extends
to large $r$, Eq.~(\ref{eq:5.2.1}) makes it obvious that
the intercept $\Delta_{\Pom}$ is controlled by the behavior of
$\sigma_{\Pom}(r)$ at $r\sim R_{c}$ and is relatively insensitive
to the large-$r$ behavior of $\sigma_{\Pom}(r)$.

The intercept
$\delta(\beta)$ will have a maximum at $\beta=0$.
Then, repeating the derivation of the Green's function
(\ref{eq:3.6}) one can easily show that at large $r,r'$
the Green's function of Eq.~(\ref{eq:5.2.1}) has the behavior
\beq
K(\xi,r.r')\propto {\exp(\Delta_{\Pom}\xi)\over \sqrt{\xi}}
\exp\left(-{\mu_{G}^{2}(r-r')^{2} \over 2\xi\delta''(0)}\right) \, ,
\label{eq:5.2.3}
\endeq
which is reminiscent of the familiar multiperipheral diffusion.
Indeed, at large $r\gg 1/\mu_{G}$ a sort of the additive quark
model is recovered, in which the (anti)quark of the dipole
developes its own perturbative gluonic cloud, and the
quark-quark scattering will be
described by the multiperipheral exchange of massive vector
mesons. The Green's function (\ref{eq:4.5}) shows the emergence
of the Regge growth of the interaction radius in this limit.
It also shows that the large-$\xi$ behavior of solutions of
our generalzed BFKL equation in this region will be similar to
solutions of the Schr\"odinger equation
\beq
\left[-{\delta''(0)\over 2\mu_{G}^{2}}
{\partial^{2} \over \partial r^{2} }
+U(r)\right]\Phi= -{\partial \over \partial \xi} \Phi
=\epsilon \Phi
\label{eq:5.2.4}
\endeq
with the potential $U(r)=-\delta(0)$ which is flat at $r> R_{c}$.

The scaling limit of Eq.~(\ref{eq:2.7}) does not have the
localized solutions, see the eigenfunctions (\ref{eq:3.2}).
An important observation is that also Eq.~(\ref{eq:5.2.1})
does not have the node-free localized solutions, which vanish at
large $r$. Indeed, if such a solution had existed, then the l.h.s.
of Eq.~(\ref{eq:5.2.1}) would have vanished at $r\rightarrow \infty$.
For the same reason, we can neglect $\sigma(r)$ and $\sigma(\rho_{2})$
in the integrand of the r.h.s. of (\ref{eq:5.2.1}) and will be left
with the positive-valued, and $r$-independent,
contribution from $\sigma(\rho_{1})$.
This shows, that (\ref{eq:5.2.2}) gives a complete set of solutions.
Consequently, the so generalized BFKL equation only has the continuous
spectrum, and the partial-wave amplitudes only have the branching
point singularity (cut) in the complex-$j$ plane.

We wish to emphasize
that none of the above conclusions changes, if $K_{1}(\mu_{G}r)$ is
changed for any function which vanishes sufficiently steeply
at large $r$, the only condition is that it must have the
short-distance $1/r$
behavior dictated by the $1/k^{2}$ ultraviolet behavior of the
perturbative gluon propagator.



\subsection{The running and freezing coupling, massless gluons}

The second interesting case is of $R_{c}\rightarrow \infty$,
i.e., the case of massless gluons, $\mu_{G}\rightarrow 0$,
but with the running coupling which freezes,
$\alpha_{S}(r)=\alpha_{S}^{(fr)}=\alpha_{S}(R_{f})$,
at finite $r \geq R_{f}$. In this case the scaling invariance
of the kernel ${\cal K}$
is restored on the infinite semiaxis $\log r > \log R_{c}$, where the
eigenfunctions $E_{+}(i\nu,\xi,r)$
are again essentially identical to the BFKL set
(\ref{eq:3.2}), the spectrum of eigenvalues will evidently
be continuous, the intercept $\Delta_{\Pom}$ will be given by the
BFKL formula (\ref{eq:3.4}) with $\alpha_{S}=\alpha_{S}^{(fr)}$
and the partial waves of the scattering amplitude will have the
cut in the $j$-plane identical to that for the original BFKL
equation. Indeed, the corresponding `Schr\"odinger' equation
(\ref{eq:5.1.6}) has the continuous spectrum starting from
$\epsilon=-\Delta(0)$, and this conclusion is evidently insensitive
to the exact shape of the potential $V(z)$ at $z\lsim z_{f}$.
Here we agree  with Ross et al. [11,14] and
disagree with Lipatov [3], who concluded
that introduction of the running coupling is by itself sufficient
for transformation of the continuous spectrum of the pomeron to
the discret one. Apparently, the origin of this conclusion is in
that Lipatov [3] restricted himself to a very special form of the
solution at large $r$, rather than using the continuum solutions
(\ref{eq:5.2.2}).

The continuation of the BFKL solutions
$(\ref{eq:3.2})$ to the region of $z<z_{f}=\log R_{f}^{2}$ poses
no problems. Let us start with the
quasiclassical situation when $\alpha^{(fr)} \ll 1$, so that
$\alpha_{S}(z)=\alpha_{S}(R_{f})
\left[1+(\beta_{0}/4\pi)\alpha_{S}(R_{f})(z_{f}-z)\right]^{-1}$
is a slow function of $z$. In this case
the slowly varying running coupling can be factored out from the
integrand of Eqs.~(\ref{eq:2.6},\ref{eq:2.7}). This suggests that
to a crude approximation one may make a substitution of the fixed
coupling $\alpha_{S}$ in Eq.~(\ref{eq:3.3}) for the running
coupling $\alpha_{S}(r)$. Consequently,
Eq.(\ref{eq:5.1.6}) takes the form of the 'Schr\"odinger'
equation for a particle with the slowly varying mass
\beq
M(z)\sim {1\over \Delta''(0)}\cdot{\alpha_{S}(R_{f})\over
\alpha_{S}(r)}
\label{eq:5.3.1}
\endeq
in the
slowly varying and monotonically decreasing in the magnitude
potential
\beq
V(z) \sim -\Delta_{\Pom}{\alpha_{S}(r)\over \alpha_{S}(R_{f})}\, ,
\label{eq:5.3.2}
\endeq
which is flat, $V(z)=-\Delta_{\Pom}$, at $ z> z_{f}$.
Evidently, the solutions
$E_{+}(i\nu,\xi,r)$ with $\Delta(i\nu) > 0$ , i.e., with the
negative energy $\epsilon$, will have the underbarrier
decrease at
$z \rightarrow -\infty$, whereas the solutions with
$\Delta(i\nu) <0$ will be continued to $z \ll z_{f}$
as plane waves. This suggests that the eigenfunction
$\sigma_{\Pom}(r)$
for the rightmost singularity will decrease at $r\rightarrow 0$
faster than the $\propto r^{1}$ solution (\ref{eq:3.7}) for the
fixed-$\alpha_{S}$ scaling BFKL regime.
The case
of large values of the frozen coupling $\alpha_{S}^{(fr)}$
must be qualitatively the same.

In section 4 we have discussed the conventional DLLA solution
of the GLDAP equation, which has a limited applicability domain.
Besides this conventional DLLA solution, in the weak coupling
limit of $\alpha_{S}(r)\ll 1$ the GLDAP equation
(\ref{eq:4.1}) has the new one-parametric family of solutions which
have the Regge behavior $\propto \exp(\Delta\xi)$. It can best be
described in terms of the density of gluons
\beq
G(\xi,r)=xg(x,r)={3\sigma(\xi,r)\over \pi^{2}r^{2}\alpha_{S}(r)}\, .
\label{eq:5.3.3}
\endeq
After factoring out $\alpha_{S}(r)$ from the kernel ${\cal K}$
which we shall justify {\sl \`a posteriori}, our generalized BFKL
equation
(\ref{eq:2.7})
takes the form
\beq
{\partial G(\xi,r) \over \partial \xi}=
{3\over 2\pi^{2}}\int d^{2}\vec{\rho}_{1}\left[
{\alpha_{S}(\rho_{1})G(\xi,\rho_{1})\over \rho_{2}^{2}}+
{\alpha_{S}(\rho_{2})G(\xi,\rho_{2})\over \rho_{1}^{2}}-
{r^{2}\alpha_{S}(r)G(\xi,r)\over \rho_{1}^{2}\rho_{2}^{2}}\right]\,.
\label{eq:5.3.4}
\endeq
{\sl \`A posteriori}, we can show that the last term $\propto
\alpha_{S}(r)G(\xi,r)$ can be neglected at $\alpha_{S}(r) \ll 1$.
Then, Eq.~(\ref{eq:5.3.5}) takes the form
\beq
{\partial G(\xi,r) \over \partial \xi}=
{4\over 3}\int_{0}^{L(R,r)}dL(R,\rho)\,\,
G(\xi,\rho)\,,
\label{eq:5.3.5}
\endeq
which has the one-parametric family of the small-$r$, large-$L(R,r)$
eigenfunctions
\beq
e_{-}(\gamma,\xi,r)=\exp[\gamma L(R,r)]\exp(\Delta\xi)=
\left[{1\over \alpha_{S}(r)}\right]^{\gamma}\exp(\Delta\xi)\, ,
\label{eq:5.3.6}
\endeq
where the exponent $\gamma$ is related to the intercept $\Delta$ by
\beq
\gamma={4\over 3\Delta} \, .
\label{eq:5.3.7}
\endeq
The corresponding dipole cross section equals
\beq
\sigma_{-}(\gamma,r)=r^{2}
\left[{1\over \alpha_{S}(r)}\right]^{\gamma-1}\exp(\Delta\xi)\, ,
\label{eq:5.3.8}
\endeq
In contrast to the conventional DLLA solution (\ref{eq:4.5}) which
summs the leading powers $[L(R,r)\xi]^{n}$, our new solutions
(\ref{eq:5.3.6},\ref{eq:5.3.8}) do manifestly
summ all powers $L(R,r)^{k}$. The
accuracy of this solution can be understood making use of
(\ref{eq:5.3.6}) in Eq.~(\ref{eq:5.3.4}): the neglected term
$\propto \alpha_{S}(r)G(\xi,r)$ gives the $\propto \alpha_{S}(r)$
correction to the solution (\ref{eq:5.3.6}).

The small-$r$ considerations alone do not fix the intercept
$\Delta$ and the exponent $\gamma$, they are determined from the
matching the solution $e_{-}(\gamma,\xi,r)$ with the large-$r$
solution $E_{+}(i\nu,\xi,r)$.



\subsection{The strong coupling freezes at $R_{f}\sim R_{c}$.}

This realistic case of the major interest is
a combination of the two previous cases. It is convenient
to start from the region of $r >R_{c},R_{f}$. The only change
from Eq.~(\ref{eq:5.2.1}) will be that the running coupling
$\alpha_{S}$ must be absorbed into the integrand
\arr
\sigma_{n+1}(r)=
{3\over \pi^{3}} \int d^{2}\vec{\rho}_{1}\,\alpha_{S}(\rho_{1})
\mu_{G}^{2}
K_{1}(\mu_{G}\rho_{1})^{2}
[\sigma_{n}(\rho_{1}) +\sigma_{n}(\rho_{2})-\sigma_{n}(r)]\, ,
\label{eq:5.4.1}
\endarr
This equation has a continuum of solutions of the form
(\ref{eq:5.2.2}) with the spectral parameter $-\infty <
\beta  < +\infty$, and its large-$r$ Green's function and
the corresponding Schr\"odinger operator will be similar to
(\ref{eq:5.2.3}) and (\ref{eq:5.2.4}), respectively. Therefore,
Eq.~(\ref{eq:5.4.1}) has the continuous spectrum and the scattering
amplitude will have a cut in the complex $j$-plane.
Continuation of the large-$r$ solutions to small $r$ is not any
different from that discussed in section 5.3. Evidently, the
intercept $\Delta_{\Pom}$ of the rightmost singularity in the
$j$-plane corresponds to the eigenvalue $\delta(\beta)$ with
$\beta=0$, $\Delta_{\Pom}=\delta(0)$,
as the corresponding eigenfunction does not have a node.
For this solution $\sigma_{\Pom}(\xi,r) = const$ vs. $r$ at
large $r$, and decreases faster than $\propto r^{1}$ at
$r\rightarrow 0$. From Eq.~(\ref{eq:5.4.1}) it is obvious, that
the intercept $\Delta_{\Pom}$ is mostly controlled by the
contribution from $r \sim R_{c}$.
Also, repeating the considerations of section 5.1,
we can exclude the node-free localized
solutions of Eq.~(\ref{eq:5.4.1})
with $\sigma(r)$ vanishing at large $r$.
This shows, that our generalized BFKL equation only has the
continuous spectrum and the
partial-wave amplitudes only have the cut
in the complex $j$-plane.



\subsection{Is the discret spectrum of the pomeron possible?}

The discret spectrum of the pomeron, i.e., generation of isolated
poles in the complex-$j$ plane, comes along with the existence of
localized solutions of the generalized BFKL equation.
The above elimination of localized solutions
is rigorous in the framework of our minimal
infrared regularization. Technically, it is based on the kernel
${\cal K}$ being finite at $r\rightarrow \infty$, see
Eqs.~(\ref{eq:5.2.1},\ref{eq:5.4.1}),
which gives rise to $\sigma(r) \sim const$ at
$r> R_{c}$ and to the continuum of solutions of the form
(\ref{eq:5.2.2}). The interaction
picture which emerges at $r> R_{c}$ has much intuitive appeal:
each well separated quark of the beam (target) dipole developes
the perturbative gluonic cloud of its own and a sort of the
additive quark model is recovered. At first sight, the localized
solutions with the vanishing perturbative
total cross section for colour dipoles of large size,
$\sigma(r)\rightarrow 0$ at $r\rightarrow \infty$,
look quite unphysical.

Here we we wish to present the qualitative
arguments in favor of a possibility of localized solutions.
The plausible scenario for the discret spectrum of the pomeron is
as follows:
The kernel ${\cal K}$ is proportional to the probability
of radiation of perturbative gluons, see Eq.~(\ref{eq:2.5}).
 When the quarks of the colour
dipole are a distance $r\gg R_{c}$ apart, the nonperturbative
colour fields stretched between the quarks may strongly modify
the vacuum and suppress the perturbative gluonic fluctuations
on the nonperturbative background in the vicinity of
quarks. Because the nonperturbative background fields are
strongly anisotropic, this suppression of the perturbative
gluonic fluctuations also will be anisotripic. However, to
a very crude approximation this suppression can
be modelled by the decrease with increasing $r$ of the
effective (nonlocal) perturbative coupling $\alpha_{S}(r,\rho)$
and/or the increase of $\mu_{G}$ with increasing $r$, resulting
in the decrease of the kernel ${\cal K}$ with increasing $r$.
In terms of the Schr\"odinger equation (\ref{eq:5.2.4}) this amounts
to a rise of the potential
$U(r)$ towards large $r$. (One can draw a useful
analogy with the asymptotic freedom decrease of
$\alpha_{S}(r)$ and the related rise of the $V(z)$
Eq.~(\ref{eq:5.3.2}) in the Schr\"odinger equation
(\ref{eq:5.1.6}).) As a natural scale at which such a suppression
of the perturbative fluctuations can take place one can think of
the confinement radius $R_{conf}$, which is of the order of the
diameter of hadrons, $R_{conf} \sim 1-2$f.
Evidently, the energy $\epsilon$ of the lowest
state will be higher than the bottom of the potential well, so
that the intercept $\Delta_{\Pom}$ of the rightmost singularity
will be lowered compared to the case of the minimal regularization.
If with the increasing $r$ the potential $U(r)$ flattens at the
still negative value $U(\infty)=-\Delta_{c}$, then the Schr\"odinger
equation (\ref{eq:4.6}) will have the continuous spectrum starting
with $\epsilon = -\Delta_{c}$ and in the complex
$j$-plane there will be a cut from $j=1+\Delta_{c}$ to $j=-\infty$
and certain number of isolated poles to the right of the cut,
at $1+\Delta_{c}< j \leq 1+\Delta_{\Pom}$.
If $U(\infty) >0$, then the cut in the complex $j$-plane will
start at $j=1$.

Although the above considerations are very qualitative ones, we
regard them as giving a sound correlation between the discret
spectrum of the pomeron and
the nonperturative suppression of the perturbative
gluonic fluctuations in large colour dipoles. (We do not
suggest a dynamical mechanism for such a suppression, though.)
In section 5.3 we explained why we
disagree with Lipatov's conclusion [3] that the running
coupling does by itself generate the discret spectrum of the
pomeron (see section 5.3).
Ross and Hancock [11] try to eliminate the contribution of
the infrared region by the $k^{2}$-cutoff in the asymmetric
form of the BFKL equation (\ref{eq:3.11}) and find a discrete
spectrum of the pomeron for few models of the nonperturbative
gluon propagator. The significance of their findings is not
clear to us, since the Hancock-Ross procedure violates the
symmetry of the cutoff of gluon propagators and its consistency
with gauge invariance is questionable (for the criticism
of this procedure see section 3). For this reason the
correspondance between the Ross-Hancock cutoff and the above
nonperturbative suppression of the perturbative gluon
fluctuations is not evident to us.



\section{The evaluation of $\Delta_{\Pom}$ and the pomeron dipole
cross section}

The analytic solution of Eqs.~(\ref{eq:2.6},\ref{eq:2.7})
at finite $R_{c}$ and with the
running coupling is not available, and we resort to the numerical
analysis. We study the large-$\xi$ behavior of
numerical solutions of Eq.~(\ref{eq:2.7}) and verify that
$\sigma(\xi,r)$ has the same asymptotic behavior
$\sigma(\xi,r)\Longrightarrow \sigma_{\Pom}(r)\exp(\Delta_{\Pom}\xi)$
(within the overall normalization)
irrespective of the boundary condition at $\xi=0$.
Namely, we compute the effective intercept
\beq
\Delta_{eff}(\xi,r)=
\partial \log\sigma(\xi,r)/\partial \xi
\label{eq:6.1}
\endeq
and check that at large
$\xi$ the effective intercept $\Delta_{eff}(\xi,r)$ tends to the
same limiting value $\Delta_{\Pom}$ for all $r$. Flattening of
$\Delta_{eff}(\xi,r)$ as a function of $r$ and as a function of
$\xi$ takes place at about the same value of $\xi$, and in all the
cases the flattening was good to a few per mille.
In Fig.1 we present the eigenfuction $\sigma_{\Pom}(r)$
for few values of the gluon correlation radius
$R_{c}=1/\mu_{G}$ (we keep the
same running strong coupling $\alpha_{S}(r)$
which freezes at $\alpha_{S}^{(fr)}=0.8$). As we have
anticipated above, $\sigma_{\Pom}(r)$ flattens at large $r$
and decreases towards small $r$ faster than the BFKL solution
(\ref{eq:3.7}), but slower that $\propto r^{2}$.

The physical interaction picture behind this
emergence of the factorizing $r$ and $\xi$ dependences
of the asymptotic cross section is as follows:
Consider, for instance, interaction of two small-sized dipoles
$r_{A},r_{B} \ll R_{c}$. Radiation of perturbative gluons by
such dipoles is suppressed by colour cancellations and by
the small coupling $\alpha_{S}(r)$. However, the perturbative
gluons start sticking out of the initially small dipoles at
a distance $r \sim R_{c}$. Consequently, after few steps of
radiation and/or after few units in $\xi$
the gluonic cloud builds up which has the size $\sim R_{c}$,
the evolution of which is controlled by $\alpha_{S}(R_{c})\approx
\alpha_{S}^{(fr)}$, and whose interaction properties do not
depend on the size of the initial beam and target dipoles.
The size of the beam and target dipoles only controls the rate
of the evolution of the limiting gluonic cloud (for the more
discussion on this point see below).

In Fig.2 we present the intercept $\Delta_{\Pom}$
for few values of $\mu_{G}=0.3,\,0.5,\,0.75,\,1.0$ GeV.
As we have emphasized above, this intercept
is controlled by interactions of the semiperturbative gluons
at transverse distances $r \sim R_{c}$, where the strong coupling
is rather large, $\alpha_{S}(R_{c})\approx  \alpha_{S}^{(fr)}$.
For the comparison, with $\alpha_{S}=0.8$ the BFKL
formula (\ref{eq:3.4}) gives $\Delta_{\Pom}(BFKL)=2.12$. We remind
that the BFKL formula is
only valid in the scaling limit of $R_{c}\rightarrow \infty$ and
obviously overestimates the intercept for the case of the finite
$R_{c}$. In the literature one often cites the
CK lower bound [9]
\beq
\Delta_{\Pom} > {3.6 \over \pi}\alpha_{S}^{(fr)}
\label{eq:6.2}
\endeq
This bound was derived in the model with the CK
sharp cutoff $k^{2}>k_{0}^{2}$
in the asymmetric form of the BFKL equation (\ref{eq:3.11})
with massless gluons, $m^{2}=0$. For a closer comparison
with the CK bound we substitute $\alpha_{S}^{(fr)}$
in Eq.~(\ref{eq:6.2}) for the
${\rm min}\left\{\alpha_{S}(\mu_{G}^{2}),0.8\right\}$. The
resulting bound is shown in Fig.2 and is not supported by our
results. The significance of the CK bound is questionable,
though, since the derivation [9] of the CK bound is plagued by
violation of the symmetry of the BFKL
equation, for a criticism see section 3.

In terms of the exponent $\gamma$ of the small-$r$ solutions
(\ref{eq:5.3.6}), we have $\gamma-1=1.19,\,1.81,\,2.33,\, 2.70$
at $\mu_{G}=0.3,\,0.5,\,0.75,\, 1.0\,$GeV, respectively. The
smaller is $\mu_{G}$, the smaller is $\gamma-1$ and the steeper
is the decrease of $\sigma_{\Pom}(r)$ at small $r$, in agreement
with the results shown in Fig.1  (for more discussion on the
relevance of solutions (\ref{eq:5.3.6}) to $\sigma_{\Pom}(r)$
see below, section 7.3).

In section 5.5 we discussed a possibility of generation of the
discret spectrum of the pomeron via the (purely hypotetical)
mechanism of the nonperturbative suppression of the perturbative
gluonic fluctuations in colour dipoles of large size. Such a
suppresion must lower the intercept of the pomeron. For a
crude evaluation of the possible effect, we introduce in
front of the kernel ${\cal K}$ the suppression factor
\beq
{\cal K}\Longrightarrow {2\over
1+\exp\left[\theta(r-R_{conf}){(r-R_{conf})^{2}/d^{2}}\right]}
{\cal K} \, .
\label{eq:6.3}
\endeq
We consider the case of $\mu_{G}=0.75$GeV, i.e., $R_{c}=0.275$f, and
take $d=0.5$f and $R_{conf}=1.5$f, i.e., impose strong suppression
of the perturbative gluonic fluctuations already at $r$ of the
order of the diameter of hadrons.
In terms of the Schr\"odinger equation
(\ref{eq:5.2.4}) such a steep suppression factor corresponds to
an abrupt rise of the potential $U(r)$ at $r\approx R_{conf}$. All
this serves to enhance the discussed effect.
We find a negligible,
less than one per cent change of $\Delta_{\Pom}$ with respect to
the case of $R_{conf}=\infty$. Such a small effect could have easily
been anticipated: the intercept $\Delta_{\Pom}$ is predominantly
sensitive to the behavior of $\sigma_{\Pom}(r)$ at $r\sim R_{c}$,
and although we have
enforced very dramatic suppression of the perturbative gluonic
fluctuations, the effect on the intercept is small because of
the strong inequality $R_{c}\ll R_{conf}$. This interpretation
is confirmed by Fig.~3, in which we show the
ratio of the $\sigma_{\Pom}(R_{conf}=1.5f,r)$ to the
$\sigma_{\Pom}(R_{conf}=\infty,r)$, and this ratio
is essentially identical to unity up to $r\sim 5R_{c}\approx 1.4$f.
This result strongly suggests, that the intercept of the pomeron
is the semiperturbative quantity, insensitive to the truly
large-distance effects.

In this paper we only concentrate on the rightmost singularity,
and we do not evaluate the gap between the poles in the discret
spectrum. We only comment, that the convergence of
$\Delta_{eff}(\xi,r)$ to the limiting value $\Delta_{\Pom}$
becomes faster when the large-$r$ cutoff (\ref{eq:6.3}) is
imposed, which suggests that the rightmost singularity is a pole
separated from other singularities by a gap.

Consequently,
with all the due reservations on the large-distance QCD, we
conclude that the intercept of the perturbative QCD pomeron is
significantly higher
than the effective intercept $\Delta_{\Pom}(hN) \sim 0.1$ as
given by the phenomenology of hadronic scattering [15].
The plausible scenario, suggested by the observed slow rise
of the hadronic cross sections, is that the exchange by perturbative
gluons is only a small part of $\sigma_{tot}(hN)$ at moderate
energies ([5], for the early discussion of this scenario see [16]).
The detailed phenomenology of the hadronic cross sections will be
presented elsewhere.



\section{The GLDAP and BFKL evolutions in
deep inelastic scattering}


\subsection{Deep inelastic scattering and the dipole cross section}

The representation (3) is universal and also applies to the deep
inelastic scattering [4,6]. The wave functions of the (T) transverse
and (L) longitudinal virtual
photon of virtuality $Q^{2}$ were derived in [6] and read
\beq
\vert\Psi_{T}(z,r)\vert^{2}={6\alpha_{em} \over (2\pi)^{2}}
\sum_{1}^{N_{f}}Z_{f}^{2}
\{[z^{2}+(1-z)^{2}]\varepsilon^{2}K_{1}(\varepsilon r)^{2}+
m_{f}^{2}K_{0}(\varepsilon r)^{2}\}\,\,,
\label{eq:7.1.1}
\endeq
\beq
\vert\Psi_{L}(z,r)\vert^{2}={6\alpha_{em} \over (2\pi)^{2}}
\sum_{1}^{N_{f}}4Z_{f}^{2}\,\,
Q^{2}\,z^{2}(1-z)^{2}K_{0}(\varepsilon r)^{2}\,\,,
\label{eq:7.1.2}
\endeq
where
\beq
\varepsilon^{2}=z(1-z)Q^{2}+m_{f}^{2}\,\,\,.
\label{eq:7.1.3}
\endeq
In Eqs.~(\ref{eq:7.1.1})-(\ref{eq:7.1.3})
$m_{f}$ is the quark mass and $z$ is the
Sudakov variable, i.e. the fraction of photon's light-cone
momentum $q_{-}$ carried by one of the quarks of the pair
($0 <z<1$). Then
\beq
\sigma_{T,L}(\gamma^{*}N,\xi,Q^{2})=
\int_{0}^{1} dz\int d^{2}\vec{r}\,\,
\vert\Psi_{T,L}(z,r)\vert^{2}\sigma(\xi,r)
\label{eq:7.1.4}
\endeq
and the structure function is calculated as
\beq
F_{2}(\xi,Q^{2})={Q^{2}\over 4\pi^{2}\alpha_{em}}
\left[\sigma_{T}(\gamma^{*}N,\xi,Q^{2})+
\sigma_{L}(\gamma^{*}N,\xi,Q^{2})\right] \, .
\label{eq:7.1.5}
\endeq
Making use of the properties of the modified  Bessel functions,
after the $z$-integration one can write
\arr
\sigma_{T}(\gamma^{*}N,\xi,Q^{2})=
\int_{0}^{1} dz\int d^{2}\vec{r}\,\,
\vert\Psi_{T}(z,r)\vert^{2}\sigma(\xi,r)\propto
{1\over Q^{2}} \int_{1/Q^{2}}^{1/m_{f}^{2}}
{d r^{2} \over r^{2} }{\sigma(\xi,r)\over r^{2}}
\label{eq:7.1.6}
\endarr
Notice, that the factor $1/Q^{2}$ in Eq.~(\ref{eq:7.1.6}),
which provides the Bjorken scaling,  comes
from the probability of having the $q\bar{q}$ fluctuation
of the highly virtual photon.
The ratio $\sigma(\xi,r)/r^{2}$ is a relatively smooth function
of $r$, which slowly rises towards small $r$, for the
conventional DLLA
behavior of $\sigma(\xi,r)$ see Eqs.~(\ref{eq:4.2},\ref{eq:4.5}),
for the weak-coupling solution of our generalized
BFKL equation see Eq.\,(\ref{eq:5.3.8}). How the dipole
cross section $\sigma(\xi,r)$ is probed in deep inelastic
scattering, will be discussed elsewhere
([17], see also [18]). Therefore, for the semiquantitative
understanding of
the transition from the GLDAP to the BFKL evolution
we can concentrate on the $r$ and $\xi$ dependence of $\sigma(\xi,r)$.



\subsection{The transition between the BFKL and DDLA regimes}

One usually discusses the low-$x$ behavior of structure functions
in terms of the DLLA solution of the GLDAP evolution equations.
In section 4 we already commented on the breaking of DLLA
at large $\xi=\log({x_{0}\over x})$. Here we wish to
concentrate on a comparison of the $\xi$-dependence of
the perturbative GLDAP and BFKL solutions for $\sigma(\xi,r)$.
The question of whether there is a strong, experimentally
observable, difference between the BFKL and GLDAP evolutions,
is being discussed in the literature for quite a time ([19]
and references therein).

The DLLA solution to GLDAP evolution
(\ref{eq:4.5}) gives the effective intercept
\beq
\Delta_{DDLA}(\xi,r)={\partial \over \partial\xi}
\left[2\sqrt{\eta}-{1\over 2}\log\xi\right]=
{4\over 3}L(R,r)\left[{1\over \sqrt{\eta}}-{1\over 2\eta}\right] \, ,
\label{eq:7.2.1}
\endeq
which must hold at $\eta \gsim 1$ and moderately large $\xi \gsim 1$.
The DLLA intercept (\ref{eq:7.2.1}) vanishes at large $\xi$. On
the other hand, in the BFKL regime the effective intercept
(\ref{eq:6.1}) tends to $\Delta_{\Pom}$ irrespective of the
radius $r$.
This suggests the practical criterion: the DLLA approximation breaks
at such values of $\xi=\xi_{c}(r)$, when
$\Delta_{DLLA}(\xi,r)$ becomes smaller than $\Delta_{\Pom}$:
\beq
\Delta_{DLLA}(\xi_{c}(r),r) \leq \Delta_{\Pom}
\label{eq:7.2.2}
\endeq
Notice, that since in expansion (\ref{eq:4.5}) the
saddle point corresponds to $n\sim \sqrt{\eta}$, this criterion
when combined with the effective intercept (\ref{eq:7.2.1})
essentially coincides with the DLLA breaking estimate
(\ref{eq:4.6}).

For a meaningfull comparison of the GLDAP and BFKL evolutions,
we must take the identical initial conditions.
For the sake of definitness we take $R_{c}=0.275$f and
consider scattering on the
proton target, so that our initial condition for Eq.~(\ref{eq:2.7})
is given by
\beq
\sigma(\xi=0,r)={3\over 2}\int d^{2}\vec{R}_{12}
|\Psi_{p}(R_{12})|^{2}\sigma_{0}(R_{12},r)\, ,
\label{eq:7.2.3}
\endeq
where $R_{12}$ is a separation of quarks "1" and "2" in the proton
(for the details see [6]).
Since the DLLA  asymptotics
(\ref{eq:4.5}) works at $\xi \gsim 1$, we shall consider the
cross section (\ref{eq:7.2.3}) as a result of the GLDAP evolution
by $\xi_{0}$ units from a lower energy, i.e., we take for the
DLLA  solution
\beq
\sigma_{DLLA}(\xi,r)=\sigma(0,r)
\sqrt{{ \xi_{0}\over\xi_{0}+\xi}}\exp\left[
2\sqrt{{4\over 3}\L(R_{f},r)}\left(\sqrt{\xi_{0}+\xi}-\sqrt{\xi_{0}}
\right)\right] \,.
\label{eq:7.2.4}
\endeq
It satisfies $\sigma_{DLLA}(\xi=0,r)=\sigma(0,r)$ by the
construction, and we use it for evaluation of the DLLA intercept
in Eq,~(\ref{eq:7.2.1}). The same $\sigma_{0}(r)$ is taken for the
boundary condition for the BFKL equation (\ref{eq:2.7}).
We assume this boundary condition to correspond to $x=x_{0}\approx
3\cdot 10^{-2}$ (the more detailed phenomenology of the deep
inelastic scattering will be presented elsewhere [17]).

Notice, that Eq.~(\ref{eq:4.2}) is the leading-logarithm formula
defines $L(R_{f},r)$ up to an additive
constant $c\lsim 1$, which depends on the detailed form of the
initial condition for ther GLDAP evolution, but which
can be neglected in the DLLA limit of
$L(R_{f},r) \gg 1$. We compute $\Delta_{eff}(\xi=0,r)$ from our
BFKL equation (\ref{eq:2.7}) and make the readjustment
\beq
L(R_{f},r) \longrightarrow \log\left[{\alpha_{S}^{(fr)}\over
\alpha_{S}(r)}\right] +c
\label{eq:7.2.5}
\endeq
such that
$\Delta_{DLLA}(\xi=0,r)$ of Eq.~(\ref{eq:7.2.1}) gives a good
approximation of $\Delta_{eff}(\xi=0,r)$ at small $r$.
With $\xi_{0}=1.25$ this is achieved taking $c\approx 0.05$.

The results of such a comparison of the BFKL and GLDAP evolutions
are shown in Fig.~4. At $\xi =\log({x_{0}\over x})\sim 1$,
the both DLLA and BFKL effective
intercepts are smaller than $\Delta_{\Pom}$ at $r\gsim 0.2$f
and are larger than $\Delta_{\Pom}$ at smaller $r$.
A good agreement of the BFKL and DLLA effective intercepts at
small $r$ is not surprizing, since our generalized BFKL equation
(\ref{eq:2.7}) has
the GLDAP equation as a limiting case at small $r$ (also see
a discussion in [5]). The $\xi$-evolution of the DLLA and BFKL
effective intercepts is very much different, though: The BFKL
effective intercept starts flattening as a function of $r$ and
tends to $\Delta_{\Pom}$, rising at large $r$
and decreasing at small $r$. In the opposite to that, the DLLA
intercept monotonically decreases with $\xi$ at all $r$, until
the GLDAP breaking (\ref{eq:7.2.2}) takes place.

The boundary between the BFKL and DLLA regimes is given by
our Eq.~(\ref{eq:4.9}). The intercept
$\Delta_{\Pom}$ is numerically small,
$\Delta_{\Pom}=0.4$ at $\mu_{G}=0.75$, and the large
numerical factor
\beq
{4\over 3\Delta_{\Pom}^{2}} \approx 8
\label{eq:7.2.6}
\endeq
emerges in the r.h.s. of Eq.~(\ref{eq:4.9}). Consequently,
the DLLA to GLDAP evolution, and the GLDAP evolution by itself,
may remain numerically viable in
quite a broad range of $\xi$, relevant to the kinematical range
of HERA.

Closer inspection of Fig.~4 reveals  a certain regularity:
$\Delta_{DLLA}(\xi,r)$ decreases with $\xi$ faster than
$\Delta_{eff}(\xi,r)$, and the rate of the divergence is higher
the larger is the radius $r$. Hence, we look at the combined
$r$ and $\xi$ dependence of $\sigma(\xi,r)$.



\subsection{Testing the DLLA identity for the gluon distribution}

In the DLLA, the gluon structure function $G(\xi,r)$ satisfies
the equation (we consider $N_{f}=3$ active flavours)
\beq
\kappa(\xi,r) = {3\over 4}\cdot {1\over G(\xi,r)}
{\partial^{2} G(\xi,r)\over \partial \xi\,\partial L(R,r)} =1\, .
\label{eq:7.3.1}
\endeq
We shall refer to the equality $\kappa(\xi,r)=1$ as the DLLA
identity. One can easily evaluate $\kappa(\xi,r)$ for the
experimentally measured gluon distributions, and
it is tempting to consider the departure from the
DLLA identity as a measure of the accuracy of the
GLDAP evolution. In this section we apply such a test of
the DLLA identity to the above described solution of
our generalized BFKL equation (\ref{eq:2.7})
subject to the boundary condition (\ref{eq:7.2.3}).

The results of such a test are shown in Fig.5.  We find that our
BFKL solution produces $\kappa(\xi,r) \approx 1$ in a very broad
range of $\xi$ and $r$ of the practical interest.
 Remarkably, the DLLA identity holds to the
$(20-30)\%$ accuracy even at large $r$, up to $r^{2} \lsim
{1\over 2}R_{c}^{2}$, and to the few per cent accuracy at
$r \lsim {1\over 3}R_{c}$. Notice somewhat oscillatory $r$-dependence
of the $\Delta_{eff}(\xi,r)$ for our BFKL solution. These
oscillations die out at large $\xi$. They have an origin in the
presence of harmonics with large $|\nu|$ and/or large $|\beta|$
in the expansion of the boundary condition (\ref{eq:7.2.3}) in
terms of eigenfunctions of the kernel ${\cal K}$ (for instance,
see Eqs.~(\ref{eq:3.5},\ref{eq:3.6}). These eigenfunctions give
the oscillating contribution to $\sigma(\xi,r)$, see
Eqs.~(\ref{eq:3.5},\ref{eq:5.2.2}). We have checked that
the DLLA identity $\kappa(\xi,r)=1$ is satisfied by
solutions of our generalized BFKL equation to a very good
accuracy in a broad range of $R_{c}$ and $\Lambda_{QCD}$.

The obvious conclusion from this observation is that the DLLA
solution (\ref{eq:4.5}), which is only valid at moderate values
of $\xi$, evolves at larger $\xi$ into our new solution
(\ref{eq:5.3.8}). These new solutions satisfy our BFKL
equation to the accuracy $\sim \alpha_{S}(r)$ and
also satisfy the DLLA identity, although it was not obvious that
necessarily $\sigma_{\Pom}(\xi,r)$ belongs to this new
family of GLDAP/BFKL solutions (\ref{eq:5.3.8}).
The exponent $\gamma$ is uniquely fixed by the pomeron intercept
$\Delta_{\Pom}$ via Eq.~(\ref{eq:5.3.7}). We have checked
that the pomeron dipole cross sections
$\sigma_{\Pom}(r)$ shown in Fig.~1 indeed to a good accuracy
satisfy the property
\beq
{\sigma_{\Pom}(r)\over r^{2}} \alpha_{S}(r)
^{\gamma-1} =const
\label{eq:7.3.2}
\endeq
up to $r\lsim {1\over 2}R_{c}$.

This observation shows that, as matter of fact, there is no real
clash between the GLDAP and BFKL evolutions if $\alpha_{S}(r)$
is running. To his end, an important notice is that the
discussion of the BFKL effects in the current literature
concentrates upon the approximation of fixed $\alpha_{S}$
(for the review and references see [20]). We have found that
the effect of the running coupling constant is quite substantial.
Firstly, the pomeron cross section (\ref{eq:5.3.8}) dramatically
differs
from the $\propto r^{1}$ BFKL scaling solution at fixed $\alpha_{S}$,
see also Fig.~1. Secondly,
one can define the counterpart of the DLLA identity
for fixed $\alpha_{S}$ too, making the substitution
\beq
L(R,r)={\beta_{0} \over 4\pi}\int_{r^{2}}^{R^{2}}
{d\rho^{2}\over \rho^{2}}\alpha_{S}(\rho)\longrightarrow
{\beta_{0} \over 4\pi}\alpha_{S}\log\left({R^{2}\over r^{2}}\right)\,.
\label{eq:7.3.3}
\endeq
Then, the BFKL scaling pomeron solution (\ref{eq:3.7}) gives
\beq
\kappa=2\log2\, ,
\label{eq:7.3.4}
\endeq
which also emphasizes a dramatic
difference between the
cases of the fixed and running strong coupling.
Therefore, matching the fixed-$\alpha_{S}$ BFKL regime at small $x$
with the GLDAP regime at larger $x$ is doomed to inconsistencies,
and such considerations are not
appropriate for the phenomenology of deep-inelastic scattering.

One interesting feature of the DLLA identity is worth while of
mention. We have calculated $\kappa(\xi,r)$ for solutions of our
generalized BFKL equation for various boundary conditions,
including the completely unphysical
 $\sigma(\xi=0,r)$ which has the form of two Gaussians peaks
with large spacing in $r$. Because of the diffusion in the radius
$r$, the dip between the two Gaussians fills in very rapidly. As
soon as such a $\sigma(\xi,r)$ becomes a relatively smooth
function of $r$, this is followed by a
rapid approach of $\kappa(\xi,r)$
to unity. We find $\lsim$10-15$\%$ departure of $\kappa(\xi,r)$
from unity
already at $\xi \gsim$2-3
and $\lsim 5\%$ departure at $\xi \gsim 5$. This fullfillment
of the DLLA identity takes place when $\sigma(\xi,r)$ has the
$r$-dependence still very different from the typical DLLA
solution and/or the asymptotic pomeron cross section
$\sigma_{\Pom}(r)$. One plausible interpretation of this
observation is that apart from the solution for the
rightmost singularity at $j=1+\Delta_{\Pom}$, the (approximate)
DLLA identity holds for other solutions in a relatively broad
range of $j \lsim 1+\Delta_{\Pom}$. The corresponding analysis
goes beyond the scope of the present paper.

We conclude this section by the prediction of the universal
scaling violation at asymptotically large $\xi$.
Namely, making use of
(\ref{eq:5.3.8}) in Eq.~(\ref{eq:7.1.6}), we obtain
\beq
F_{2p}(x,Q^{2})\propto \left[{1\over \alpha_{S}(\hat{q}^{2})}
\right]^{{4\over 3\Delta_{\Pom}}}
\left({1\over x}\right)^{\Delta_{\Pom}}
\label{eq:7.3.5}
\endeq
(The possible difference between $Q^{2}$ of the virtual photon
and the scale $\hat{q}^{2}$ in the strong coupling in the r.h.s of
this formula, and the conditions for the onset of this universal,
$x$-independent, scaling violation will be discussed elsewhere [17].)
The results of Fig.~4 suggest, though, that the onset
of this universal scaling violation is well beyond the
kinematical range of the HERA experiments.



\section{The beam-target symmetry and the boundary condition}

In the above discussion we have suppressed the target size
variables $r_{B}$ but, evidently,
it is the generalized energy-dependent dipole-dipole cross section
$\sigma(\xi,\vec{r}_{A},\vec{r}_{B})$ which
emerges as the fundamental
quantitiy of the lightcone $s$-channel approach to the diffractive
scattering. The lowest-order dipole-dipole cross section has an
obvious beam-target symmetry property
$\sigma_{0}(\vec{r}_{A},\vec{r}_{B})=
\sigma_{0}(\vec{r}_{B},\vec{r}_{A})$. This beam-target symmetry is
but a requirement of the Lorentz-invariance and one has to
have it extended to all energies. Thus, the beam-terget symmetry
emerges as an important constraint on the admissible boundary
condition for the BFKL equation.
In the derivation of our generalized
BFKL equation we have treated the $s$-channel gluon $g_{s}$ of
Fig.~6 as belonging to the beam-dipole $A$. The gluon-induced
correction to the total cross section was reinterpreted in
terms of the generalized dipole cross section
$\sigma(\xi,\vec{r}_{A},\vec{r}_{B})$, see Eq.~(\ref{eq:2.5}).
Alternatively, we could have treated the same $s$-channel gluon
$g_{s}$ as belonging to the target-dipole, and the result must
have been the same. The two descriptions differ in that in the
former the perturbative $t$-cahnnel gluons $g_{1}$ and $g_{1'}$
enter the kernel ${\cal K_{A}}$, which we supply with the
subscript $A$ as it acts on the beam variable $r_{A}$ of the
dipole-dipole cross section $\sigma(\xi,r_{A},r_{B})$,
whereas the gluons $g_{2}$ and $g_{2'}$ are the exchanged gluons
in the dipole-dipole cross section (\ref{eq:1.1}). In the latter
description $g_{2},g_{2'}$ enter the kernel ${\cal K_{B}}$ which
now acts on the target variable $r_{B}$ of the dipole-dipole
cross section $\sigma(\xi,\vec{r}_{A},\vec{r}_{B})$, and
$g_{1},g_{1'}$ will become the exchanged gluons in the input
dipole-dipole cross section. The beam-target symmetry constraint
essentially implies, that the boundary condition for the BFKL
evolution must be calculable in the same perturbation theory as
the one used to construct the generalized BFKL kernel ${\cal K}$
Eq.~(\ref{eq:2.6}). (To this end we remind, that the $K_{1}(x)$
in the kernel ${\cal K}$ is precisely the derivative
of the gluon propagator (correlation function) $K_{0}(x)$.) This
kernel-cross section
relationship is crucial for having the beam-target symmetry, which
will be violated with the arbitrary choice of the boundary
condition for the dipole-dipole cross section
$\sigma(\xi=0,\vec{r}_{A},\vec{r}_{B})$.



\section{Restoration of the factorization at asymptotic energies}

It is well known that in the conventional Regge theory the
isolated poles in the complex-$j$ plane give rise to the factorizing
scattering amplitudes. Even if the factorization holds for separate
singularities in the $j$-plane, the scattering amplitude which is
a sum of contributions from many close-by singularities, does not
factorize. For instance, the lowest order dipole-dipole cross section
(\ref{eq:1.1}) manifestly breaks the factorization relation.
It is interesting to notice, that the factorization of the total
cross sections restores at asymptotic energies.

Indeed, for the rightmost singularity in the $j$-plane
the beam-target symmetric dipole-dipole total cross section will
have the factorized form
\beq
\sigma_{\Pom}(\xi,\vec{r}_{A},\vec{r}_{B})=
\sigma_{\Pom}(r_{A})\sigma_{\Pom}(r_{B})\exp(\xi\Delta_{\Pom})\, .
\label{eq:9.1}
\endeq
By virtue of Eq.~(\ref{eq:1.2}) this implies that
at asymptotic energies
\beq
\sigma_{tot}(AB)\propto
\left[\int dz_{A} d^{2}\vec{r}_{A}
|\Psi(z_{A},\vec{r}_{A})|^{2}
\sigma_{\Pom}(r_{A})\right]\cdot
\left[\int dz_{B} d^{2}\vec{r}_{B}
|\Psi(z_{B},\vec{r}_{B})|^{2}
\sigma_{\Pom}(r_{B})\right]\cdot
\exp(\Delta_{\Pom}\xi) \, ,
\label{eq:9.2}
\endeq
which has the manifestly factorized form. From the point of view
of the phenomenology of the total cross sections, this
factorization
will be broken by the unitarization corrections needed to tame
too rapid a rise of the perturbative bare pomeron cross section.
The above predicted universal scaling violation in deep inelastic
scattering is a particular case of this factorization restoration.



\section{Measuring the BFKL pomeron at HERA}

The experimental determination of the pomeron intercept
$\Delta_{\Pom}$ requires isolation of the BFKL pomeron cross
section, which is not an easy task because of the close
similarity of the BFKL and GLDAP solutions, discusssed in
detail in Section 7.
The GLDAP evolution is a special limiting case of the generalized
BFKL evolution.
Every and each solution of the BFKL equation does not satisfy the
GLDAP evolution and the DLLA identity. Our remarkable finding is
that the small-$r$ pomeron cross section (\ref{eq:5.3.8}) which
is a solution of the BFKL equation, satisfies the GLDAP
evolution equation within the correction terms $\sim \alpha_{S}(r)$.
Consequently, one may expect that
GLDAP evolution with the properly
posed boundary condition can be a good approximation to the BFKL
evolution. The GLDAP considerations can not by themselves fix this
boundary condition, though. Inspection of
Fig.4 shows that the DLLA and BFKL effective intercepts are close
to each other and to the
$\Delta_{\Pom}$ in a broad range of $x$ of the
practical interest at
\beq
r\approx r_{0}= (0.4-0.5)R_{c}\,.
\label{eq:10.1}
\endeq
Consequently, choosing boundary condition $G(\xi,r_{0})\propto
x^{-\Delta_{\Pom}}$ at $r\approx r_{0}$,
one will obtain the GLDAP solutions, which at $r \lsim r_{0}$
for all the practical
purposes will be indistinguishable from the BFKL solution.
Eq.(\ref{eq:7.1.6})
shows that the proton structure function receives contributions
from the broad range of $r$, in which the effective intercept
$\Delta_{eff}(r)$ is substantially different from $\Delta_{\Pom}$.
For this reason studying the $x$-dependence of $F_{2}(x,Q^{2})$
is not suitable for accurate determination of the pomeron
intercept.

Fig.4 suggests
that zooming at $r\approx$0.15-0.2f one can determine
the pomeron intercept $\Delta_{\Pom}$ already at the moderately
large ${1\over x}$ of the HERA experiments. This can be achieved
measuring either the longitudinal structure function
$F_{L}(x,Q^{2})$ or the scaling violations in the transverse
structure function $\partial F_{T}(x,Q^{2})/\partial \log(Q^{2})$.
Making use of the wave functions (\ref{eq:7.1.1},\ref{eq:7.1.2})
one can easily show that these quantities are local probes of
$\sigma(\xi,r)$ at
\beq
r^{2}\approx {B\over Q^{2}} \,  ,
\label{eq:10.2}
\endeq
where for the longitudinal structure function $B_{L}\approx 12$
and $B_{T}\approx 5$ for the transverse structure function.
Consequently, experimentally $\Delta_{\Pom}$ can best be
estimated as
\beq
\Delta_{\Pom}=-{\partial \log F_{L}(x,Q^{2}) \over \partial
\log x}
\label{eq:10.3}
\endeq
at $Q^{2}=Q_{L}^{2}\approx $12-25\,GeV$^{2}$ and/or
\beq
\Delta_{\Pom}=-{1\over  F_{T}(x,Q^{2})}\cdot
{\partial^{2} \log F_{T}(x,Q^{2}) \over \partial
\log x \,\partial\log Q^{2} }
\label{eq:10.4}
\endeq
at $Q^{2}=Q_{T}^{2}\approx $5-8\,GeV$^{2}$. Our prediction is
that these derivatives must be $x$-independent already starting
with $x\lsim 10^{-2}$. The above  difference by
the factor $\approx (2.5-3)$ between $Q_{L}^{2}$ and $Q_{T}^{2}$
is an important consequence of wave functions
(\ref{eq:7.1.1},\ref{eq:7.1.2}). Purely kinematically, the smaller
$Q^{2}$ implies the broader range of ${1\over x}$, and
for this reason the second
determination (\ref{eq:10.4}) may prove more accurate one.
A comparison of the two
determinations (\ref{eq:10.3}) and (\ref{eq:10.4}) is an
important consistency check.
We conclude that from the point of view of measuring the
pomeron intercept, the HERA experiments must concentrate on
accurate measurements of $F_{2}(x,Q^{2})$ and $F_{L}(x,Q^{2})$
in the region of moderate $Q^{2}\lsim 30$ GeV$^{2}$.

\section{Conclusions}

The purpose of this paper has been to understand the spectrum of
eigenvalues of the generalized BFKL equation [4,5] for the dipole
total cross section. Our emphasis was on
the realistic case of the finite correlation radius for the
perturbative gluons and of the freezing strong coupling.
The advantage of our generalized BFKL
equation (\ref{eq:2.6},\ref{eq:2.7}) is an
easy introduction of the finite gluon correlation radius in the
manner which is consistent with the gauge invariance constraints.

We have shown that our generalized BFKL equation
(\ref{eq:2.6},\ref{eq:2.7}) has the continuous spectrum, which
corresponds to the QCD pomeron described by the cut in the
complex $j$-plane. We have determined a dependence on the size
$r$ of the dipole for the rightmost singularity in the $j$-plane
and found the corresponding intercept $\Delta_{\Pom}$.
It is much smaller than given by the BFKL formula (\ref{eq:3.4})
and even smaller than the lower bound cited in [9], but
is substantially larger than the
phenomenological value $\Delta_{\Pom}(hN) \sim 0.1$.
we have investigated a transition from the GLDAP to the BFKL
evolution in the limit of large ${1\over x}$, and have shown
that the two equations have an overlapping asymptotic solution
in the weak coupling limit. The phenomenologically important
implication is that the GLDAP evolution with the correctly posed
boundary condition remains a viable description of deep inelastic
scattering well beyond the kinematical range of the HERA experiments.
Nonetheless, the pomeron intercept can be determined already
from the HERA data on deep inlastic scattering, and
we have suggested practical methods of measuring $\Delta_{\Pom}$.
\bigskip\\
{\bf Acknowledgements}: B.G.Z. and V.R.Z. are grateful to
J.Speth for the hospitality at IKP, KFA J\"ulich.
\pagebreak

\newpage
{\bf Figure captions:}

\begin{itemize}
\item[Fig.1 - ]
The pomeron dipole cross section $\sigma_{\Pom}(r)$ for different
values of $\mu_{G}$. The straight lines show the $r^{1}$ and $r^{2}$
behavior.

\item[Fig.2 - ]
The intercept $\Delta_{\Pom}$ for $\mu_{G}=0.3,\,0.5,\,0.75,\,1.0\,$
GeV (shown by triangles). The solid curve shows
the Collins-Kwiecinski
 lower bound (\ref{eq:6.2}).

\item[Fig.3 - ]
The ratio of the pomeron dipole cross section with the cutoff
(\ref{eq:6.3}) of the large-$r$ controbution to the pomeron
dipole cross section without the cutoff. The cutoff radius
$R_{conf}=1.5$f, the gluon correlation radius $R_{c}=0.275$f.

\item[Fig.4 - ]
Comparison of effective intercepts of the DLLA solution
(\ref{eq:7.2.4}) to the GLDAP evolution and of the solution of
our BFKL equation. Both solutions start with the identical dipole
cross section at $x=3\cdot 10^{-2}$. The pomeron intercept
$\Delta_{\Pom}=0.4$ is shown by the horizontal line.

\item[Fig.5 - ]
Test of the DLLA identity for the solution of our generalized
BFKL equation. The gluon correlation radius $R_{c}=0.275$f.

\item[Fig.6 - ]
One of diagrams of the driving term of the rising dipole cross
section.
\end{itemize}
\end{document}